# Dual Universality and Unconventional Phase Diagram of a Clean 2D Superconductor with Tunable Disorders

Puhua Wan[1], Qihong Chen[1,2], Oleksandr Zheliuk[1], Le Zhang[3,1],

Minpeng Liang[1], Xiaoli Peng[1], Jianting Ye[1]*

[1] *Device Physics of Complex Materials, Zernike Institute for Advanced Materials, University of Groningen, 9747 AG Groningen, The Netherlands.*
[2] *Beijing National Laboratory for Condensed Matter Physics, Institute of Physics, Chinese Academy of Sciences, Beijing 100190, China.*
[3] *Shenzhen Key Laboratory of Laser Engineering, College of Physics and Optoelectronic Engineering, Shenzhen University, Shenzhen 518060, China.*

*\* e-mail: j.ye@rug.nl*

**Weakly disordered two-dimensional (2D) superconductors can host richer quantum phase transitions than their highly-disordered counterparts[1–6]. This is due to the insertion of a metallic state in the transition between a superconductor and an insulator. Disorders were predicted to show profound influences on the boundaries surrounding this intermediate metallic phase, affecting the existence of a metallic ground state called Bose metal[1,2] and the dynamic processes in a quantum Griffiths state[7,8]. Here we present a study on quantum phase transitions of a clean 2D superconductor, $MoS_2$, under a perpendicular magnetic field as a function of disorder strength that is tuned electrostatically. We found two universal scaling behaviors independent of disorders: a power-law scaling applies for all metallic states, and an activated dynamical scaling characterizes transitions between quantum Griffiths state and weakly localized metal. The phase diagram in this unexplored clean regime shows that a Bose metal ground state is expected only below a critical disorder. Whereas, stronger disorders can stabilize a true 2D superconductivity and enhances the quantum Griffiths phase. These behaviors are diametrically different from the conventional understanding of the clean regime.**

The disorder has significant influences on macroscopic quantum coherent states of superconductors. Physically, the presence of disorders requires treating the disorder scattering prior to superconductivity, where the electrons in the normal state are described by a diffusive motion. The mean free path $l_m$ of the electron, which is determined by the strength of disorder scattering, becomes an important parameter that defines the so-called "cleanness" of a superconductor. Compared with coherence length $\xi_0$ found in the superconducting state, $l_m \sim \xi_0$ marks the borderline towards the clean ($l_m \gg \xi_0$) or dirty ($l_m \ll \xi_0$) limits. In two-dimensional (2D) systems, the increase of disorder can drastically suppress the superconductivity by reducing the transition temperature $T_c$ and increasing the normal state resistance $R_N$. When the disorder reaches a critical strength in many granular films, these highly-disordered 2D superconductors ($l_m \ll \xi_0$) turn into insulators at a universal resistance close to $h/4e^2$ [9,10]. Transport properties measured in the vicinity of the transitions often exhibit a universal scaling behavior, belonging to the universal class of dirty bosons[11–13].

On the other hand, two distinct features were observed in many weakly-disordered systems ($l_m \sim \xi_0$) based on 2D single crystals[6,14,15] and granular thin films with enhanced crystallinity[16,17]. First of all, an intermediate metallic state emerges after suppressing the superconductivity in a magnetic field $B$, which was theoretically proposed to be a quantum metallic phase. At zero temperature, the ground state of this metallic state was predicted to be a Bose metal[1–4]. As described in the Bose metal model, the superconductor to quantum metal transition (SMT) originates from the activation of free vortex dislocations following a power-law dependence on the $B$ field. This gives rise to a finite resistance even if the field is infinitesimal[1–3,14]. Secondly, a quantum Griffiths state, in which the Cooper pairs are isolated in superconducting rare regions under a large $B$ field, dominates the phase transitions from a quantum metal to a weakly-localized metal (normal-metal state)[5,6]. The transition dynamics are increasingly frozen when temperature decreases. At zero temperature, the dynamical exponent $zv$ becomes infinite reaching the so-called quantum Griffith singularity[8,18,19]. Compared with the constant dynamical rate found in dirty Bosons, here, the $zv$ value varies as a function of temperature at the phase boundary between the quantum Griffiths and the weakly-localized metallic states.

Although the phase diagram in the dirty regime has been studied extensively[11,20,21], the clean regime ($l_m > \xi_0$) remains largely unexplored, where disorders are predicted to influence both quantum metal and quantum Griffith state as a function of external magnetic field $B$ applied perpendicular to the 2D plane. It was theoretically proposed that depending on the interplay between disorder strength and quantum fluctuation at zero temperature[1,2], in the low $B$ field regime, the critical field $B_{c0}$ –the external magnetic field required for the system to enter the metallic state– can be either $B_{c0} = 0$ or $B_{c0} > 0$. This corresponds to a proposed ground state of either a Bose metal or the zero-resistance state of a true 2D superconductor, respectively. Experimentally, although having different disorders can lead to zero or non-zero $B_{c0}$ values for different ground states, it is an open question whether the intermediate metallic states all belong to a universality class of power-law scaling. In the large $B$ field regime, where the superconducting rare regions dominate the phase transitions, a quantum Griffiths state is theoretically predicted for all systems with non-zero disorders[7,8]. The activated dynamic scaling has been widely studied in the quantum Griffiths states of many 2D superconductors[6,17,22,23] at individual states of static disorders. Whereas, studying the universality requires preparing a range of states, where the scaling behavior can be analyzed for different disorders in searching for identical behaviors. It is encouraging that a recent preprint reported that the activated dynamical scaling is universal in the dirty regime by varying the film thickness[24]. Nevertheless, it remains unknown whether the rare region effect will also dominate the clean regime since the spatial fluctuation of the order parameter decreases when the disorder is weakened[25]. Therefore, it is also highly demanded to find a clean system with variable disorders to exam the validity of the proposed universal exponent $\psi\nu$[7,8,26,27].

Compared with the extensive studies on universal scaling in the dirty regime[11,20,21,24], the research in the clean regime ($l_m > \xi_0$) is seriously limited by the available 2D systems where scaling analyses can be performed. As shown in Fig. 1a, the clean regime of 2D superconductors is mostly accessed by bulk 3D single crystals with strong anisotropy, which behaves as 2D superconductors (the third quadrant, lower left). The granular films and atomically thin superconducting single crystals are affected by the intrinsic defects and being air unstable[28,29]. Except for monolayer $NbSe_2$ protected by $h$-BN[14,30], other films and interfaces all reside on the dirty side (first quadrant, upper-right). To resolve these difficulties, ionic gating on intrinsically semiconducting 2D crystals stands out as the choice, which can induce 2D superconductivity on air-stable single-crystalline channels[31,32], hence achieving weak disorders[15]. Nevertheless, the conventional ionic gating[33] is still hampered by the low mobility, therefore, remains at the first quadrant.

To prepare a range of states in the second quadrant of Fig. 1a, we developed the following ionic gating with enhanced mobility. This starts with preparing ion-gated $MoS_2$ transistors as shown in Fig. 1b and c. The Hall bar channel is shaped by a 50 nm thick $Al_2O_3$ film that isolates the field effect of ionic liquid. All carriers are induced by a single ionic gating and fixed by freezing the device below the glass transition temperature $T_g$ of the ionic liquid. This initial gating (up to 6 V) is also limited to the electrostatic regime, where identical states can be reproduced in two consecutive gatings (Fig. S1). As shown in the simplified Helmholtz double layer formed on the channel (Fig. 1d), strong gating pushed the ions close to the channel surface (upper panel) generating non-uniform potentials, acting as disorders. In the cases of monolayer channels, these disorders can be strong enough to completely localize the electrical transport, reentering highly insulating states[34,35]. On the other hand, the discreteness of ion can be smoothed out by thermally releasing the ions away from the channel, forming a more uniform potential when the double layer thickness $z$ becomes larger (lower panel of Fig. 1d). Therefore, the strength of the disorder can be tuned electrostatically by varying the distance $z$ between the ions and the induced carriers.

By grounding the ionic gate and warming the device up to a temperature slightly higher than $T_g$, disorder states with different strengths can be accessed through sequential thermal releases. It worth noting that the thermally activated ionic motion is limited to the *z-direction* out of the channel plane. Whereas, the in-plane motion of ions is restrained due to the rubbery state of high viscosity. As a result, different disorder strengths can be accessed in a single device with identical disorder distribution in the *xy* plane. Compared with different disorders accessed by preparing multiple samples, either by thin film deposition or post annealing[20,36], varying disorder strength in the same device also maintains the identical intrinsic disorders caused by structural defects and impurities.

The disorder strength can be parameterized as $W \sim l_m^{-1/2}$ [37,38] (section 1 of the supplementary materials). From the band structure calculation of the ion-gated $MoS_2$ [39], the carriers are primarily doped on the outmost layer at the *K/K'* points of the Brillouin zone. Since transport measurement can determine the 2D carrier density $n_{2D}$ (by Hall effect, Fig. S2) and sheet resistance $R_s$ measured right before superconducting transition, we can obtain $l_m = v_F \tau$, where Fermi velocity $v_F$ and scattering time $\tau$ are calculated in section 1 of the supplementary materials. As shown in Fig. 1f, we prepared *in situ* 11 different states on a single device (device A). The effect of having different disorders is conspicuous because the $R_s$ varies oppositely to the change of $n_{2D}$ when both parameters are measured at a normal state ($T$ = 13 K) above the superconducting transition. With the decrease of $n_{2D}$, the $R_s$ conversely decrease as well due to a faster increase of Hall mobility $\mu_H$ (derived in Eq. 5 of supplementary materials) from 119 to 834 cm$^2$V$^{-1}$s$^{-1}$ (inset of Fig. 1e). Therefore, the $l_m$ values increase from 12.4 to 71.6 nm when the disorder strength reduces. The first state, having a $T_c$ = 7.1 K (with $l_m$ = 12.4 nm) is induced by the initial gating at the right side of the superconducting dome of MoS$_2$ induced by field effect[32]. The $T_c$ is defined at the temperature when transition reaches half of the normal resistance $R_N$ (measured at $T$ = 13 K). By thermal releasing, both $\mu_H$, and $l_m$ follow the variation of the superconducting dome as a function of the $n_{2D}$ (inset of Fig. 1e).

When a magnetic field is applied to a highly-disordered 2D superconductor, the superconducting islands gradually lose the long-range correlations. At the zero-temperature limit, transition appears at the quantum resistance, $h/4e^2$, where the islands are isolated by insulating regions[25]. In the vortex tunneling model, the magnetoresistance of such a system varies exponentially with the $B$ field as log$R \sim B$ [40–42]. In contrast, even for the most disordered state ($l_m$ = 12.4 nm shown in Fig. 1f), the $R_N$ of our sample –after suppressing the superconductivity at a high field– is one order of magnitude smaller than $h/4e^2$. Unlike the dome-like variation of $T_c$, the upper critical field $B_{c2}$ (derived from Fig. S3a by extrapolating to $T$ = 0 K) decreases monotonically when the system is tuned to a cleaner state with larger $l_m$ (inset of Fig. 1e). Compared with the coherence length $\xi_0$ extracted from $B_{c2} = \Phi_0/2\pi\xi_0^2$, where $\Phi_0$ is the magnetic flux quantum, the mean-free path $l_m$ is a few times larger than $\xi_0$ (Fig. S3). Therefore, our system is located inside the clean regime, where $l_m > \xi_0$. Both analyses above suggest that our states are in the weakly disordered regime, which is beyond the description of the dirty boson model[20]. Consistently, the magnetoresistances of all states, with different mean-free paths, cannot be fitted with the vortex tunneling model in the framework of dirty Bosons (Fig. S4).

Instead, the magnetoresistances can be well described by the Bose metal model[2], where $R(B)$ follows the power-law scaling $R \sim (B - B_{c0})^{2\nu_{SMT}}$. Here, $B_{c0}$ is the critical field that can cause the transition from a zero-resistance state to a metallic state (SMT), $\nu_{SMT}$ the exponent of the superfluid correlation length. By fitting the magnetoresistance isotherm (shown in Fig. 2a) with power-law scaling as a function of external $B$ field for different disordered states, we can obtain the $B_{c0}$ and $\nu_{SMT}$ as the fitting parameters. As shown in Fig. 2a, we plot $R_s$ as a function of $\Delta B = B - B_{c0}$ for different disordered states measured at $T$ = 1.9 K. The gray dashed lines show the best fitting using the power-law scaling. For a representative state with $l_m$ = 28.4 nm (Fig. 2b), the magnetoresistances at different temperatures can also be well described by the power-law scaling yielding the temperature dependence of $\nu_{SMT}$. In the present system with a wide range of disorders in the clean regime, the power-law scaling shows universal validity. The varying exponent $\nu_{SMT}$ from each disordered state are plotted in Fig. 2c as a function of temperature. Note the $\nu_{SMT}$ values are all obtained below $T_Q$, a temperature below which the exponentially diminishing thermal effect is overwhelmed by the quantum fluctuations (section 3 of supplementary materials). The dashed line at $\nu_{SMT} = 0.5$ (Fig. 2c) indicates the limit of unhindered flux flow, where the vortex-pinning effect is absent. Consequently, vortexes can move freely, and the $R_s$ shows a linear dependence on the $B$ field: $R_s \sim B/B_{c2}$ [14,43,44]. Crossing the SMT boundary, the superfluid correlation length scales as $\xi_+ \sim (B - B_{c0})^{-\nu_{SMT}}$ in the quantum-metal phase[2]. Since the $\nu_{SMT} \geq 0.5$, the state with larger $\nu_{SMT}$ shows a slower divergence in $\xi_+$ with the increase of the magnetic field. Following each disorder state shown in Fig. 2c, an increase in temperature enhances the dynamics, causing a decrease of $\nu_{SMT}$ due to the enhanced thermal fluctuation (section 3 of supplementary materials). For a vertical cut in Fig. 2c at $T = 1.9$ K, when the system is thermally released towards the cleaner states, the $\nu_{SMT}$ for different disorder states decreases and approaches the limit $\nu_{SMT} = 0.5$ for the cleanest state with $l_m = 65.2$ nm. At higher temperatures ($T = 2.5, 3.6,$ and $4.4$ K, *etc.*), where the thermal fluctuation is stronger, unhindered vortex flow can be accessed by states with larger disorders ($l_m = 59.5, 51.1,$ and $44.3$ nm, *etc.*).

The excitation current can also assist the vortex motion via Lorentz force[43,44]. In another measurement (device B), a zero-resistance state below the superconducting transition can be observed only by applying a small excitation current (Fig. S5). The leveling of $R_s$, which has been interpreted as the evidence of quantum metal at $T = 0$ [14,15,41], appears when current excitations are applied. With a small excitation (0.4 µA) in the measurement of device A, approaching $\nu_{SMT} = 0.5$ (Fig. 2c) is assisted by the thermal fluctuations at finite temperature, alongside the quantum fluctuation. In the $T \to 0$ limit, where the thermal fluctuation is absent, the quantum fluctuation is expected to destroy the phase coherence in a critical-clean 2D superconductor, forming Bose metal ground state[1]. Compared with the smooth decrease of $\nu_{SMT}$ as a function of temperature shown in clean states with large $l_m$ values, it is noted in Fig. 2c that a signature of saturation in the temperature dependence of $\nu_{SMT}$ showing in the state with $l_m = 12.4$ nm below $T = 3$ K. Such a saturation is also observed in another crystalline 2D superconductor[14].

As shown in Fig. 3, we then move to a higher magnetic field to explore the superconducting rare regions, where nearly isolated superconductivity islands are embraced by normal metals[6,22,45]. In panel A-C, we show the magnetoresistances of three representative states with high, medium, and low disorder strengths (the other states are shown in Fig. S7). Multiple crossing points can be identified in panels A and B for the states with large and medium disorders, forming a phase boundary in the $B$–$T$ phase diagram[6,45]. For the case of weak disorder strength shown in panel C, the magnetoresistances tend to saturate and collapse together at a high magnetic field. We can divide the multiple crossing points into narrow temperature ranges[45] (Fig. S8a-h). Within each division, the magnetoresistances can be analyzed by the finite-size scaling: $R(B,T) = R_c f\left(\frac{|B-B_c|}{(T/T_L)^{1/zv}}\right)$, where $R_c$ and $B_c$ are the critical resistance and critical magnetic field defined by the crossing point of the adjacent $R_s(B)$ curves, and $f$ is an arbitrary function with $f(0) = 1$, $z$ the dynamical critical exponent, $v$ the correlation length exponent, $T_L$ the lowest temperature of the division. When our system is released to cleaner states with $l_m > 44.3$ nm, determining the crossing points and extract accurate exponent $zv$ becomes increasingly difficult because the magnetoresistance curves merge (Fig. 3c).

As shown in Fig. 3d, the magnetic field dependence of $zv$ increases from a plateau $zv \approx 0.5$ for all disordered states. In a clean (2+1)D $XY$ superconductor (the extra dimension is imaginary time), $z$ is general set as 1 due to the long-range correlation[5,45], and $v = 0.5$ [5,6,17,18]. Here, the plateau at $zv \approx 0.5$ (Fig. 3d) corresponds to the clean states free from the rare region effects, which is caused by thermal fluctuation at finite temperature. Above the plateau, superconducting rare regions can be locally ordered. In the present 2D system ($d = 2$), the Harris criterion $dv > 2$ is violated when $v = 0.5$. Therefore, $z$ is expected to diverge when the temperature approaches zero [5]. Starting from the initial state with the highest disorder strength ($l_m = 12.4$ nm, purple dots), $zv$ increases from the plateau and grows rapidly towards a zero-temperature crossing point $B_c^* = 8.5$ T (extrapolated from Fig. S10). A similar trend can be observed in less disordered states, where the fast increase of $zv$ occurs at a lower $B_c^*$. Above the plateau $zv \approx 0.5$ (Fig. 3d), the phase transition is increasingly affected by the rare region effect. In the vicinity of $B_c^*$, the optimal rare regions (ORR) become exponentially rare to find and dominate the phase transition. The size of ORR, $L_{ORR}$, can be obtained from $B_c^* \sim \Phi_0/L_{ORR}^2$, where $\Phi_0$ is the magnetic flux quantum[45,46]. Consistently, as disorder strength decreases towards cleaner states in Fig. 3d, the $L_{ORR}$ increases accordingly, hence showing smaller $B_c^*$.

In contrast to the strong influence of disorders on the magnetic field dependences shown in Fig 3D, the temperature dependence of $z\nu$ shows a very similar trend for all disorder states (Fig. 3e). We analyzed the magnetoresistances following the activated dynamical scaling: $R\left(B, \ln\frac{T_0}{T}\right) = \phi\left[\delta\left(\ln\frac{T_0}{T}\right)^{\frac{1}{\psi\nu}}\right]$. Here, $T_0$ is a fitting parameter, $\psi$ tunneling exponent, $\nu$ the correlation length exponent, $\delta = |B - B_c|/B_c$ characterize the difference from the crossing point[17,47]. At finite temperature, we can obtain the $(\psi\nu)_{\text{eff}}$ from $\frac{1}{z\nu} = \left(\frac{1}{\psi\nu}\right)_{\text{eff}} \frac{1}{\ln(T_0/T)}$ by two-parameter fitting, which approaches $\psi\nu$ at zero temperature[17] (section 4 of the supplementary materials). As shown in Fig. 3e, the scaling analysis collapses independent of disorders, yielding $(\psi\nu)_{\text{eff}} = 0.68$. This is in good agreement with the theoretically predicted $\psi\nu = 0.6$ for a critical point with infinite randomness[19,48,49]. Unlike the multiple curves found in $z\nu(B)$, the collapse of $z\nu(T)$, for different strengths of disorders, suggests a universal scenario for weakly disordered 2D superconductors. Theoretically, a weak disorder can always enhance to be a strong disorder after renormalization. Thus, phase transitions with any finite disorder are governed by a critical point of infinite randomness with a universal exponent $\psi\nu$ [7,8,26,27]. To further exam the universality, we apply the same scaling to several similar 2D superconductors with weak disorders. This includes MBE-grown Ga thin films[6], annealed single-crystalline InO$_x$ films[17], and LaAlO$_3$/SrTiO$_3$ interfaces[50]. As shown in Fig. 3e, the extracted $z\nu$ can be well fitted with $(\psi\nu)_{\text{eff}}$ equals 0.61, 0.62, and 0.64, where $T_0 = 0.38$, 1.21, and 4.3 K, respectively. In addition to MoS$_2$, the consistent $(\psi\nu)_{\text{eff}}$ values also found in different materials strongly support a general universality of activated dynamical scaling in the weakly-disordered 2D superconductors.

As shown in Fig. 3e, the universality is valid when the rare region effect exists, corresponding to $T < T_c^0$. Here, $T_c^0$ is the critical temperature above which the system enters the clean state free from rare region effects. Therefore, we can obtain $T_c^0$ from $z\nu(T_c^0) = 0.5$. When $T > T_c^0 = 5.5$ K, a deviation was observed from activated dynamical scaling for the data enclosed by the dashed box (Fig. 3e), which corresponds to the plateau of $z\nu \approx 0.5$ shown in Fig. 3d. When $T < T_c^0$, rare regions can be locally ordered, dominating the scaling behavior[18]. The $T_c^0$ extracted from activated dynamical scaling is also in good agreement with that (5.8 K) derived from the Werthamer-Helfand-Hohenberg (WHH) analysis (Fig. S10) for the onset of rare region effect.

For different disorders, the $L_{\text{ORR}}$ varies with the disorder's strength. Each disorder state has a $L_{\text{ORR}}$ that corresponds to a specific $B_c$ value (Fig. 3d). However, the $T_0 = 10.4$ K, obtained from fitting $\frac{1}{zv} = \left(\frac{1}{\psi v}\right)_{\text{eff}} \frac{1}{\ln(T_0/T)}$ and extrapolating to $zv(T_0) = 0$, turns out to be identical for all accessed states in gated MoS$_2$. As a result, an identical critical temperature $T_c^0$ is obtained for states with different disorder strengths. In the rare region regime, the probability $\omega$ of finding a superconducting rare region scales with the rare region size $L_{\text{RR}}$ as a Gaussian distribution $\omega(L_{\text{RR}}) \sim \exp(-pL_{\text{RR}}^d)$, where $p$ is the disorder concentration, $d$ the dimensionality of the rare region[18]. Since the thermal release of gating only varies the disorder strength while keeping the $p$ constant, we expect the same distribution of rare regions in the vicinity of $T_c^0$. Consequently, when $\left(\frac{1}{\psi v}\right)_{\text{eff}}$ is universally valid, the fitting parameter $T_0$ is expected to be identical for different disordered states, as long as the disorder concentration holds constant. In comparison, the consistency in $T_0$ values cannot be observed the $\beta$-W thin films, where the disorder is tuned by varying the thickness of thin films in separate samples[24].

Based on the scaling analysis on MoS$_2$ in the clean regime, we could propose a phase diagram of the 2D superconductor as a function of the disorder strength in the clean regime. The phase diagram can be then compared with that proposed 30 years ago by Fisher[11], which is still widely used as general guidance for the quantum phases in 2D superconductors[10,13,20,51,52]. The magnetoresistance isotherms measured at a finite temperature corresponds to the shaded slice shown in Fig. 4a. The Fisher diagram predicts that the 2D superconductivity persists though out the clean regime up to the clean limit.

For the variable disorder states with $l_m < \xi_0$, at finite $T = 1.9$ K, the magnetoresistances measured for different disorders are plotted as a function of applied $B$ field and disorder strength, which scales as $\sim l_m^{-1/2}$. With the increase of $B$ field, quantum phases appear at both low and high field regimes, and the transitions occur between zero-resistance, quantum metal, quantum Griffiths (rare region state), and weakly-localized metal (normal metal state). In the low field regime of the phase diagram, a universal power-law scaling applies to all metallic states of different disorders, showing different $v_{\text{SMT}}$ and $B_{c0}$ values. With the reduction of disorder strengths, zero-resistance states gradually disappear due to the weaker pinning effect from the disorders. The extracted $B_{c0}$ decreases from 1.1 T for the most disordered state ($l_m = 12.4$ nm) to nearly zero for the critical state with $l_m^c = 31.7$ nm. Above this critical disorder, the vortex motion is collectively pinned. The large pinning energy might not be overcome by the quantum fluctuation when disorder strength is sufficiently high[53]. As a result, a zero-resistance can exist in states with strong disorders ($l_m < 31.7$ nm). The vortexes can be unpinned also by the enhanced quantum fluctuation[2], which is fulfilled when a magnetic field $B > B_{c0}$ is applied. Whereas, in cleaner states having $l_m > l_m^c$, free vortex dislocation can be generated by applying an infinitesimal $B$ field, forming a quantum metal. It worth noting that present samples are measured at finite temperature and finite current excitation (Fig. S5). Nevertheless, we can expect that the true 2D superconducting regime would expand if we lower the temperature further towards zero. Therefore, the Bose metal ground state (at $T = 0$) is expected in even cleaner states below the critical disorder ($l_m > l_m^c$).

In the high field regime above the quantum metal states, where $B_{RR} < B < B_c$, the quantum Griffiths states appear when the locally ordered rare regions are embraced by normal metallic regions. Here, $B_{RR}$ is extrapolated from Fig. S10, marking the $B$ field, above which the rare region effect starts to develop[6]. Considering the fast increase of $zv$ found in Fig. 3d, the quantum Griffith transition corresponds to a quick frozen of dynamics toward zero temperature, which is governed by infinite randomness. Since disorder enhances the spatial fluctuation of order parameter[25], the distribution of rare regions can then be extended to a wider $B$ field range. Therefore, the quantum Griffiths state ($B_{RR} < B < B_c$) expands with the increase of disorder in the phase diagram. Towards the clean side, the $L_{ORR}$ grows when the disorder strength decreases. Theoretically, an infinitesimal amount of disorder can already cause infinite randomness[7,8], hence, the rare-region regime is expected to close only in the ideal state with zero disorder. The order parameter of the superconducting rare regions is gradually suppressed by the increase of the $B$ field up to $B_c$ ($T = 1.9$ K), at which the system eventually enters the normal state as a weakly localized metal. Along the phase boundary (blue line in Fig. 4), the transitions from the quantum Griffiths to the weakly localized metal in the vicinity of $B_c$ ($T = 1.9$ K) can be described by an activated dynamical scaling function with a universal exponent $(\psi v)_{eff}$ (Fig. 3e).

In conclusion, we've presented a clean 2D superconductor with *in-situ* tunable disorders. The quantum phase transitions in the unexplored clean regime are characterized by two universal scaling behaviors in a perpendicular magnetic field. In the low field, the superconductor-quantum metal transitions follow a power-law scaling. While, in the high field, the quantum metal-normal metal transition is characterized by an activated power-law scaling. Besides the dual universality, a clean 2D superconductor shows an unconventional phase diagram. The true zero-resistance states appear only at a critical disorder and the 2D superconductivity enhances by the stronger disorders. Below the critical disorder, the system enters the quantum metal state at $B = 0$; where the quantum Griffiths region diminishes gradually toward the clean limit. These behaviors observed in the clean regime are strikingly different from the widely accepted Fisher diagram. Our phase diagram can act as a guide to both clean and dirty limits. To the clean limit, our result indicates that a super-clean superconductor should have a ground state of Bose metal that is also affected by infinite randomness from even infinitesimal defects. Namely, it is intriguing to study whether the Griffith regime will eventually close if we can go closer to the clean limit. On the other hand, following our phase diagram to the dirty side, variable disorders can also be prepared by strong ionic gating[34,35]. The variable strong disorder could be used to bridge the activated power-law scaling with multiple $zv$ values and the single $zv$ scaling observed towards the limit of critical disorder $\Delta_c$ (Fig. 4a, the Fisher diagram), where the process of crossover in scaling behaviors remains unknown.

**Methods**

$MoS_2$ thin flakes are mechanically exfoliated from a bulk single crystal onto the silicon substrate with an oxide layer of ~285 nm thick. The electrodes (50 nm Au on 1 nm Ti) and an $Al_2O_3$ isolation layer (50 nm) are deposited by electron beam evaporation after standard electron-beam lithography processes. Three SR830 DSP lock-in amplifiers were used for 4-probe measurement and a Keithley 2450 DC meter was used for applying gate bias $V_g$. The data presented in this work were collected from 2 devices. The 11 different disordered states were achieved on a single device (device A). Device A (channel width = 3 μm) was measured with a constant AC current of 0.4 μA. And the data shown in Fig. S5 were collected from device B (channel width = 3 μm) with various current excitations.

The first state with the highest disorder strength (smallest mean free path $l_m$) is prepared at 220 K by applying a gate voltage $V_g = 6$ V through an ionic liquid: N, N-Diethyl-N-methyl-N-(2-methoxyethyl) ammonium bis(trifluoromethanesulfonyl)imide (DEME-TSFI). The applied $V_g$ breaks the equilibrium distribution of ions and accumulates excess cations on the channel surface. This then induces free carriers on the channel by field-effect, forming the so-called Helmholtz double layer. The device is then cooled down with applied $V_g$ to 170 K (below the glass transition temperature $T_g \approx 190$ K of DEME-TSFI) to freeze the ionic motion. The $V_g$ is then removed for the remaining experiments by grounding the ionic gate electrode.

The following states with decreasing disorder strengths are subsequently obtained by warming the device to a temperature slightly higher than $T_g$. With a grounded gate, the accumulated ions tend to diffuse away from the $MoS_2$ surface towards the equilibrium state. Since applying $V_g$ breaks the equilibrium of ion distribution only along the z-direction, the thermal release of the accumulated ions is also along the same direction to larger z values. Therefore, a smoother potential on the channel surface can be obtained after each thermal release corresponding to the different disorder strengths. Due to the very slow ionic motion near the $T_g$, new states can be then controllably accessed by freezing the ionic motion again by fast cooling down. The disorder strength stays constant as long as the ions are frozen below the $T_g$.

This disorder-tuning method can ensure a few crucial technical details that substantiate the whole analysis. First of all, the disorder is variable only for its strength. The 2D disorder distribution is kept for all states due to the lack of ionic diffusion in the *xy* plane because the ions are in a very viscous state close to $T_g$. Moreover, all states are accessed after one ionic gating on the same device, which maintains the intrinsic disorder and eliminates any possible difference in chemical details accessed by two different gatings. In reality, the sample is chemically inert if gated at 220 K. Identical states can be achieved by applying the same $V_g$ (Fig. S1).

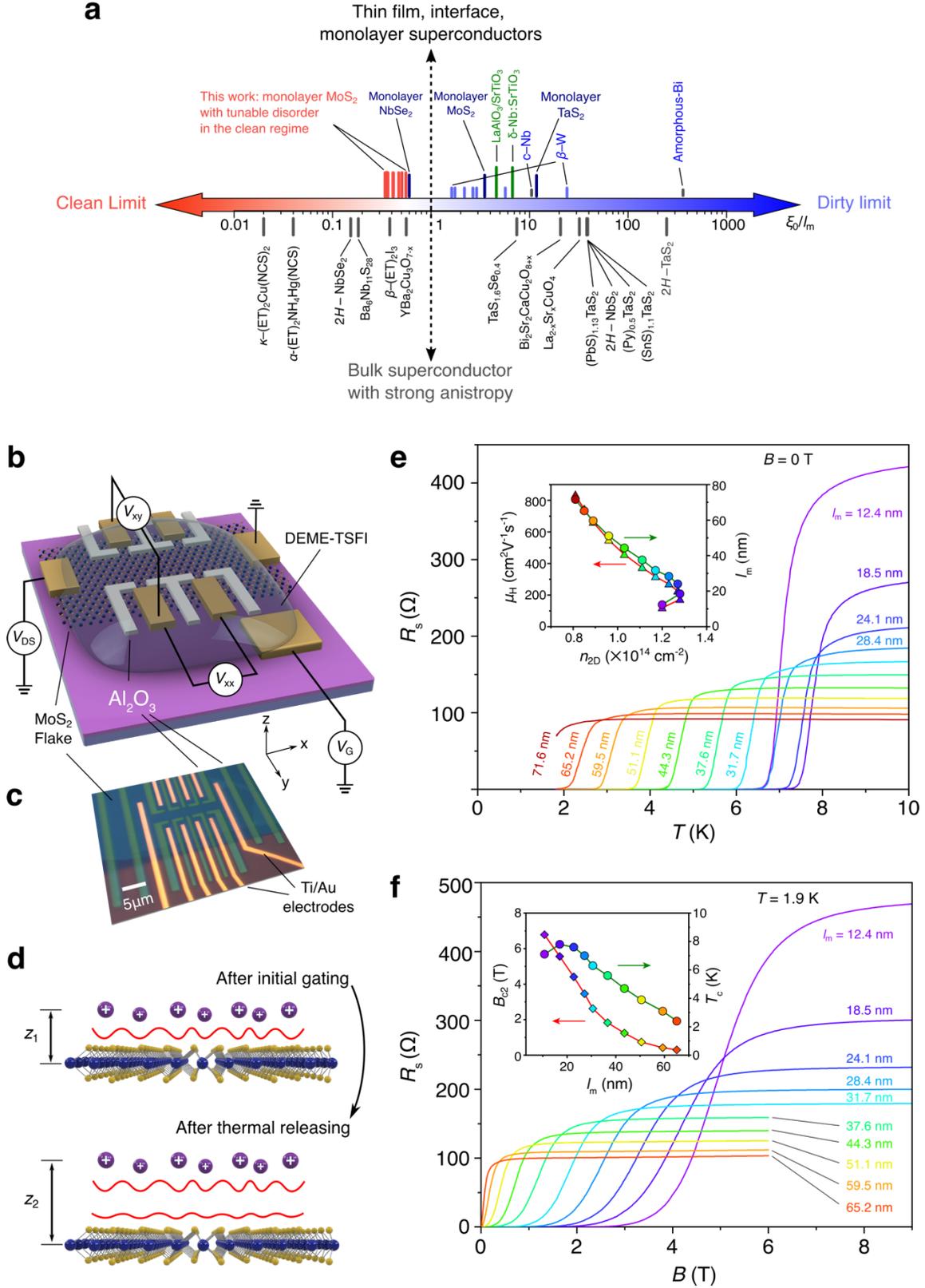

**Fig. 1 | Thermal releasing of the ionic gating for different disorder states. a,** Quadrant chart of 2D superconductors in the clean and dirty regimes evaluated as the ratio between coherent length/mean-free-path $\xi_0/l_m$. The ratios of most bulk superconductors are calculated in Pippard coherent length $\xi_P$[54]. The $\xi_P \sim \xi_0$ in pure superconductors without considering the scattering[42]. A range of disorder states is prepared in the dirty regime for $\beta-$W, and in the clean regime for

MoS$_2$, where the universality of quantum phase transitions can be analyzed. **b,** Schematics of an ion-gated MoS$_2$ device. A Hall bar structure is electrostatically defined. The conducting path is shaped to a specific geometry by coating a 50 nm thick Al$_2$O$_3$ film, which isolates the MoS$_2$ surface from ionic liquid, hence eliminates the field-effect from ions. **c,** An optical image of a typical Hall bar device on a multilayer (~ 5 nm) MoS$_2$ flake. The length of the scale bar is 5 μm. **d,** A schematic illustration of tuning the disorder strength. The initial strong gating forms a narrow Helmholtz double layer[31] with width $z_1$. By warming the frozen ionic liquid to $T > T_g$, the accumulated ions diffuse away from the channel to a larger distance $z_2 > z_1$, achieving a smoother electric field distribution on the channel and a weaker disorder strength. **e,** The temperature dependence of sheet resistance $R_s$ of different disorder states prepared *in situ* on a single device (device A, channel width = 3 μm). Inset is the Hall mobility $\mu_H$ and the mean free path $l_m$ of each state for different $n_{2D}$ values. The localization of field-induced carriers can be observed at the low mobility side, where both $l_m$ and $n_{2D}$ decrease, after passing the superconducting dome peak[32,34]. The cleanest state can reach Hall mobility $\mu_H$ ~800 cm$^2$V$^{-1}$s$^{-1}$ and a mean free path $l_m$ ~70 nm. **f,** Magnetoresistances of various disordered states at $T = 1.9$ K. Inset: upper critical field $B_{c2}$ (Fig. S3) and $T_c$ as a function of $l_m$. A dome-like dependence of $T_c$ was observed as a function of $l_m$, which is consistent with the monotonic dependence between $l_m$ and $n_{2D}$. On the other hand, the upper critical field $B_{c2}$ varies monotonically with the change of $l_m$.

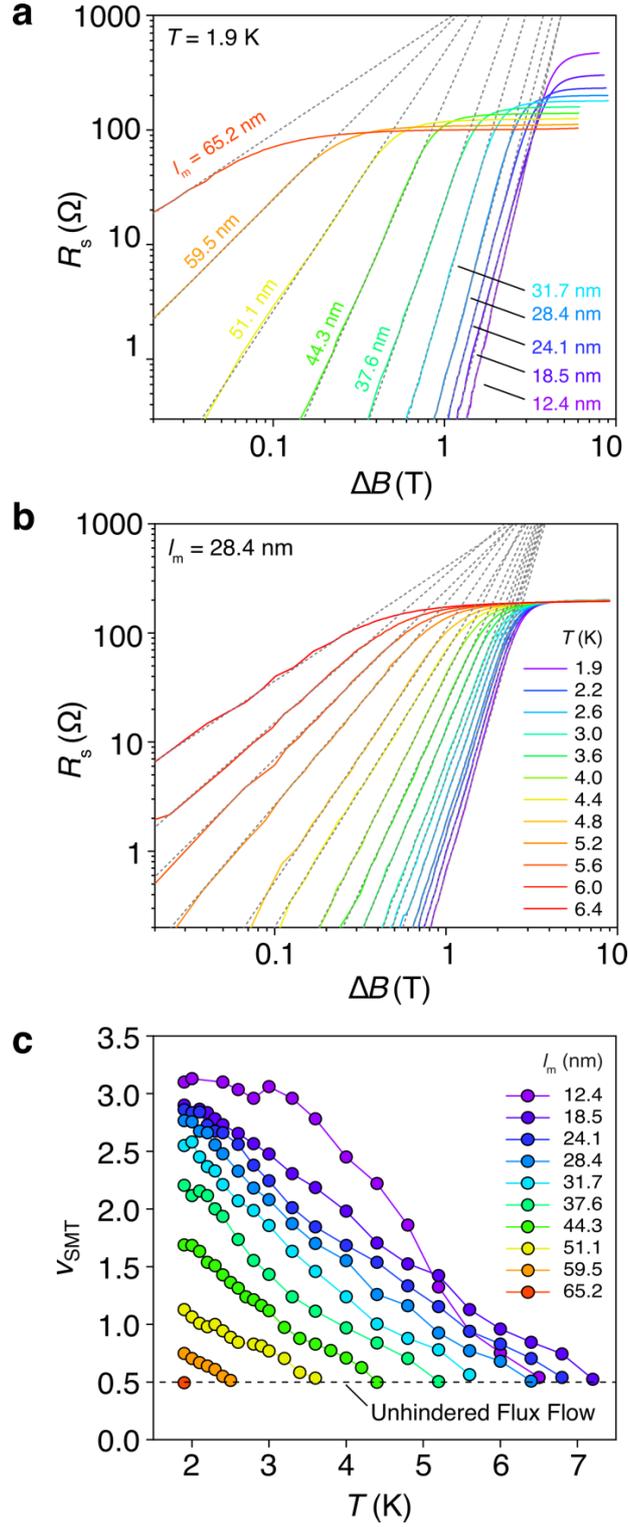

**Fig. 2 | Power-law scaling at the low magnetic field regime for different disordered states.** The magnetoresistances are fitted to $R_s \sim \Delta B^{2\nu_{SMT}}$, where $\Delta B = B - B_{c0}$, for **a,** various disorder states at $T = 1.9$ K and **b,** a representative state with $l_m = 28.4$ nm at different temperatures. The $\nu_{SMT}$ and $B_{c0}$ are tunable fitting parameters. All analyses are made below the thermal activation temperature $T_Q$ (Fig. S6). The dashed lines are the best linear fittings. **c,** The temperature dependence of the superfluid correlation exponent $\nu_{SMT}$ extracted from the magnetoresistances of states with different $l_m$ values.

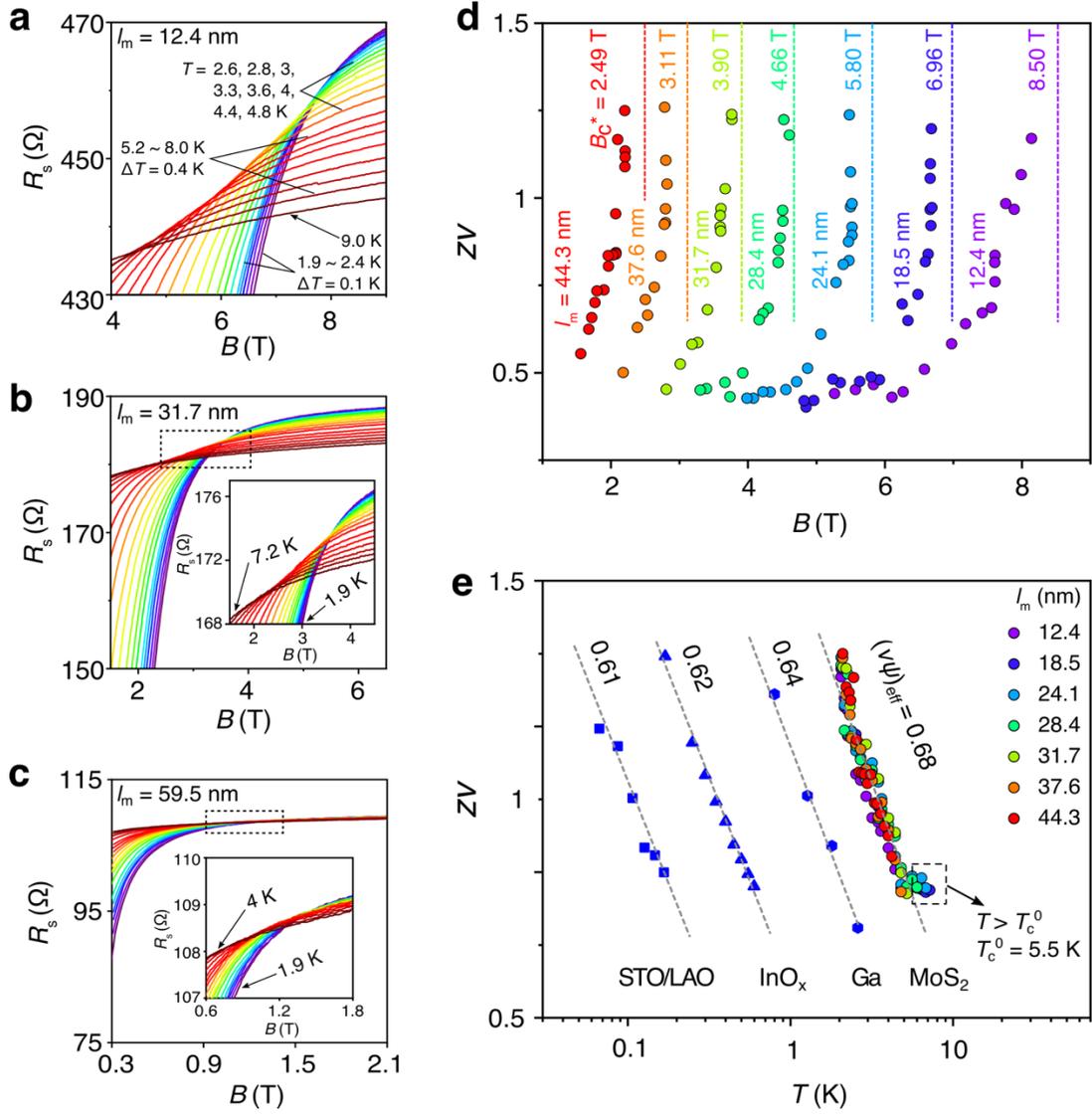

**Fig. 3 | Multiple crossing points in the magnetoresistance isotherms and the activated dynamical scaling. a-c,** Crossing points of three representative states with $l_m$ = 12.4, 31.7, and 59.5 nm, respectively. In panel a, the magnetoresistance isotherms are measured at the labeled temperatures. The temperatures in panels b and c are labeled in detail in Fig. S7. The magnetoresistance isotherms gradually collapse when the system becomes cleaner. Insets in panels B and C are the expanded area marked by the dashed squares for a closer look at the diminishing crossing regions in cleaner states. The resistance range (y-axis) is set to be identical for panels A to C. **d,** The $B$-field dependences of the exponent $zv$. For all accessed states, the $zv$ values show fast increases when approaching zero temperature crossing point $B_c^*$ (dashed line, extrapolated from Fig. S10). In the zero-temperature limit, $zv \to \infty$ at $B_c^{*8,18,19}$. **e,** The temperature dependence of $zv$ shows identical activated dynamical scaling behavior for all disordered states accessed in MoS$_2$. Similar scaling can be found in other weakly-disorder 2D superconductors with static disorders. This suggests that a universal scenario applies to the weakly disordered superconductors in this rare-region regime. The dashed box encloses the data points that deviate from the fitting. This corresponds to the high-temperature region $T > T_c^0$, where the system starts to be free from the rare region effect due to the thermal fluctuation.

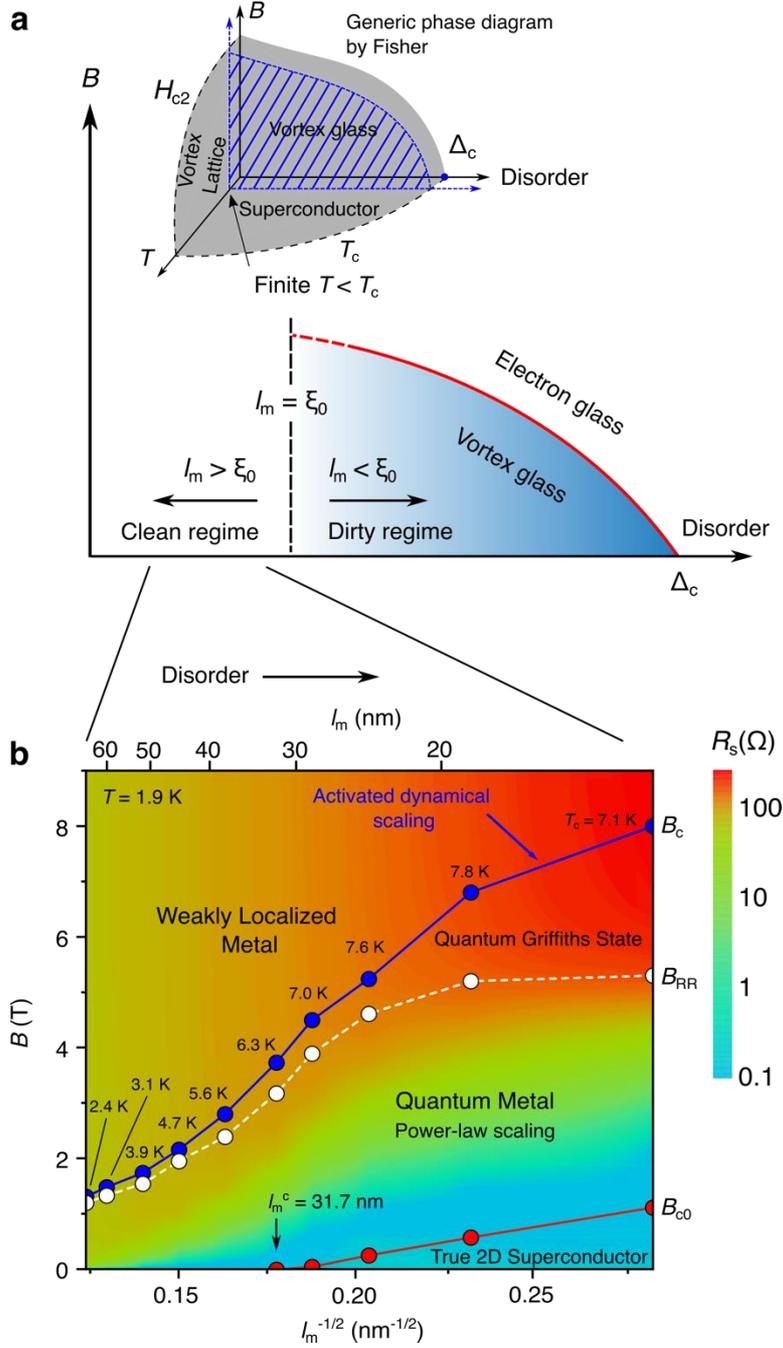

**Fig. 4 | Phase diagram measured at $T = 1.9$ K as a function of magnetic field $B$ and disorder strength that scales as $W \sim l_m^{-1/2}$. a**, Schematic phase diagram proposed by Fisher[11] as the generic guidance to the quantum phases in 2D superconductors. All experiments reported so far are confined in the dirty regime ($l_m < \xi_0$). **b**, The present measurement enters the clean regime ($l_m > \xi_0$). The individual states are marked by the corresponding $T_c$. The most disordered state has a $T_c = 7.1$ K, which is on the right side of the superconducting dome. The second state with $T_c = 7.8$ K is close to the dome peak. The $B_{c0}$ (red dots) is the boundary between a true 2D superconductor and quantum metal, following the Bose metal analysis. The $B_{RR}$ values (white dots) are extrapolated from the fitting shown in Fig. S8. The white dashed line marks the $B$ field range, from which the rare region (RR) effect starts to influence. For $B > B_{RR}$, superconducting rare regions can be locally ordered, forming a quantum Griffiths state. The $B_c$ values (blue dots) are the crossing points of two adjacent magnetoresistance isotherms

measured at $T = 1.9$ and 2 K, separating the weakly localized metal (WLM) and quantum Griffiths state. A true 2D superconductor exists only when the disorder strength is sufficiently large. When the disorder strength is reduced, the true 2D superconducting region gradually shrinks and eventually disappears at a critical mean free path $l_m^c$. Although quantum phases at zero field can be either a true 2D superconductor or a quantum metal, depending on the disorder strength, both phases follow a universal power-law scaling independent of the disorders. In the high field regime between $B_{RR}$ and $B_c$, a quantum Griffiths state exists for all disordered states. Along the blue line, in the vicinity of $B_c$, the transitions from the quantum Griffiths to the weakly localized metal follow an activated dynamical scaling function with a universal exponent $(\psi\nu)_{\text{eff}} = 0.68$.